\RequirePackage{ifpdf}
\ifpdf 
\documentclass[pdftex]{sigma}
\else
\documentclass{sigma}
\fi

\begin{document}

\allowdisplaybreaks

\renewcommand{\PaperNumber}{027}

\FirstPageHeading

\renewcommand{\thefootnote}{$\star$}

\ShortArticleName{Group Analysis of the 3D Equations of Fluids
with Internal Inertia}

\ArticleName{Applications of Group Analysis\\ to the
Three-Dimensional Equations\\ of Fluids with Internal
Inertia\footnote{This paper is a contribution to the Proceedings
of the Seventh International Conference ``Symmetry in Nonlinear
Mathematical Physics'' (June 24--30, 2007, Kyiv, Ukraine). The
full collection is available at
\href{http://www.emis.de/journals/SIGMA/symmetry2007.html}{http://www.emis.de/journals/SIGMA/symmetry2007.html}}}

\Author{Piyanuch SIRIWAT and Sergey V. MELESHKO}

\AuthorNameForHeading{S.V. Meleshko and P. Siriwat}

\Address{School of Mathematics, Suranaree University of
Technology,\\ Nakhon Ratchasima, 30000, Thailand}

\Email{\href{mailto:fonluang@yahoo.com}{fonluang@yahoo.com},
\href{mailto:sergey@math.sut.ac.th}{sergey@math.sut.ac.th}}

\ArticleDates{Received October 31, 2007, in f\/inal form February
12, 2008; Published online February 24, 2008}

\Abstract{Group classif\/ication of the three-dimensional
equations describing f\/lows of f\/luids with internal inertia,
where the potential function $W= W(\rho,\dot{\rho})$, is
presented. The given equations include such models as the
non-linear one-velocity model of a bubbly f\/luid with
incompressible liquid phase at small volume concentration of gas
bubbles, and the dispersive shallow water model. These models are
obtained for special types of the function $W(\rho,\dot{\rho})$.
Group classif\/ication separates out the function
$W(\rho,\dot{\rho})$ at 15 dif\/ferent cases. Another part of the
manuscript is devoted to one class of partially invariant
solutions. This solution is constructed on the base of all
rotations.  In the gas dynamics such class of solutions is called
the Ovsyannikov vortex. Group classif\/ication of the system of
equations for invariant functions is obtained. Complete analysis
of invariant solutions for the special type of a~potential
function is given.}


\Classification{76M60; 35Q35}

\renewcommand{\thefootnote}{\arabic{footnote}}
\setcounter{footnote}{0}

\section{Introduction}

The article focuses on group classif\/ication of a class of
dispersive
models \cite{bk:GavrilyukTeshukov2001}\footnote{See also references therein.}
\begin{gather}
\dot \rho +\rho  {\rm div} (u)=0,\qquad \rho \dot u+\nabla p=0,\nonumber \\
p=\rho\frac{\delta W}{\delta \rho }-W=\rho \left( \frac{\partial
W}{\partial \rho }-\frac \partial {\partial t}\left(
\frac{\partial W}{\partial \dot \rho }\right) -{\rm div} \left(
\frac{\partial W}{\partial \dot \rho }u\right) \right)
-W,\label{eq:main}
\end{gather}
where $t$ is time, $\nabla $ is the gradient operator with respect
to the
space variables, $\rho $ is the f\/luid density, $u$ is the velocity f\/ield, $%
W(\rho ,\dot \rho )$ is a given potential, ``dot'' denotes the
material time
derivative: $\dot f=\frac{df}{dt}=f_t+u\nabla f$, and $\frac{\delta W}{%
\delta \rho }$ denotes the variational derivative of $W$ with respect to $%
\rho $ at a f\/ixed value of $u$. These models include the
non-linear one-velocity model of a bubbly f\/luid (with
incompressible liquid phase) at small volume concentration of gas
bubbles (Iordanski~\cite{bk:Iordanski[1960]},
Kogarko~\cite{bk:Kogarko[1961]},
Wijngaarden~\cite{bk:Wijngaarden[1968]}), and the dispersive
shallow water model (Green \& Naghdi~\cite{bk:GreenNaghdi[1976]},
Salmon~\cite{bk:Salmon[1998]}). For the Green--Naghdi model, the
potential function is \cite{bk:GavrilyukTeshukov2001}
\[
W(\rho,\dot\rho)=\rho(3g\rho-\varepsilon^2%
\dot\rho^2)/6,
\]
 where $g$ is the gravity, $\varepsilon$ is the ratio of the
vertical length scale to the horizontal length scale. For the
Iordanski--Kogarko--Wijngaarden model, the potential function is
\cite{bk:GavrilyukTeshukov2001}
\[
W(\rho ,\dot \rho )=\rho \big( c_2\rho _{20}\varepsilon _{20}
(\rho_{20})-2\pi n\rho _{10}R^3\dot R^2\big) ,
\]
where
\[
\frac 43\pi nR^3=\left( \frac 1\rho -\frac{c_1}{\rho _{10}}\right)
, \qquad \rho_{20}=c_2\left( \frac 1\rho -\frac{c_1}{\rho
_{10}}\right) ^{-1},
\]
$\varepsilon _{20}$ is the internal energy of the gas phase, $c_1$
and $c_2$ are the mass concentrations of the liquid and gas
phases, $n$ is the number of bubbles per unit mass, $\rho _{10}$
and $\rho _{20}$ are the physical densities of components. The
quantities $c_1$, $c_2$, $n$ and $\rho _{10}$ are assumed
constant.

One of the methods for studying of dif\/ferential equations is
group analysis \cite{bk:Ovsiannikov[1978]}. Many applications of
group analysis to partial dif\/ferential equations are collected
in \cite {bk:Handbook}. Group analysis beside construction of
exact solutions provides a regular procedure for mathematical
modeling by classifying dif\/ferential equations with respect to
arbitrary elements. An application of group analysis involves
several steps. The f\/irst step is the group classif\/ication with
respect to arbitrary elements. This paper considers group
classif\/ication of equations (\ref{eq:main}) in the
three-dimensional case, where
the function  $W_{\dot \rho \dot \rho }$ satisf\/ies the condition $%
W_{\dot \rho \dot \rho }\neq 0$. Notice that for
$W_{\dot\rho\dot\rho}=0$ or
$W(\rho,\dot\rho)=\dot\rho\varphi(\rho)+ \psi(\rho)$, the momentum
equation becomes
\[
\dot u+\psi^{\prime\prime}\rho_x=0.
\]
Hence in the case $W_{\dot\rho\dot\rho}=0$, equations
(\ref{eq:main}) are similar to the gas dynamics equations. This
case has been completely studied \cite{bk:Ovsiannikov[1994]} (see
also~\cite{bk:Fragment401}).

The one-dimensional case of equations (\ref{eq:main}) was studied
in \cite{bk:HematulinMeleshkoGavrilyuk[2007]}. As in the case of
the gas dynamics equations there are dif\/ferences in the group
classif\/ications of one-dimensional and three-dimensional
equations.

Another part of this paper is devoted to a special vortex
solution. This solution was introduced by L.V.~Ovsyannikov
\cite{bk:Ovsiannikov[1995]} for ideal compressible and
incompressible f\/luids. This is a partially invariant solution,
generated by the Lie group of all rotations. L.V.~Ovsyannikov
called it a ``singular vortex''. It is related with a special
choice of non-invariant function. He also gave complete analysis
of the overdetermined system corresponding to this type of
partially invariant solutions: all invariant functions satisfy the
well-def\/ined system of partial dif\/ferential equations with two
independent variables. The main features of the f\/luid f\/low,
governed by the obtained solution, were pointed out in
\cite{bk:Ovsiannikov[1995]}. It was shown that trajectories of
particles are f\/lat curves in three-dimensional space. The
position and orientation of the plane, which contains the
trajectory, depends on the particle's initial location. Later
particular solutions of the system of partial dif\/ferential
equations for invariant functions were studied in
\cite{bk:Popovych[2000],bk:Chupakhin[2003],bk:CherevkoChupakhin[2004],bk:Pavlenko[2005]}.
For some other models, this type of partially invariant solutions
was considered in
\cite{bk:HematulinMeleshko[2002],bk:Golovin[2005]}. Exact
solutions in f\/luid dynamics generated by a rotation group are of
great interest by virtue of their high symmetry. The classical
spherically symmetric solutions is one of the particular cases of
such solutions.

In this manuscript a singular vortex of the mathematical model of
f\/luids with internal inertia is studied. Complete group
classif\/ication of the system of equations for invariant
functions is given. All invariant solutions for this system are
presented.

\section{Equivalence Lie group}

Since the function $W$ depends on the derivatives of the dependent
variables, for the sake of simplicity of f\/inding the equivalence
Lie group, new dependent variables are introduced:
\[
u_5=\dot \rho ,\qquad \phi _1=W,\qquad \phi _2=W_\rho ,\qquad \phi
_3=W_{\dot \rho },
\]
where $u_4=\rho $ and $x_4=t$. An inf\/initesimal operator $X^e$
of the equivalence Lie group is sought for in the form
\cite{bk:Meleshko[2005]}:
\[
X^e=\xi ^i\partial _{x_i}+\zeta ^{u_j}\partial _{u_j}+\zeta ^{\phi
_k}\partial _{\phi _k},
\]
where all coef\/f\/icients $\xi ^i$, $\zeta ^{u_j}$ and $\zeta
^{\phi _k}$ ($i=1,2$, $j=1,2,3,4,5$, $k=1,2,3$) are functions of
the variables\footnote{In the classical approach \cite[Chapter~2,
Section~6.4]{bk:Ovsiannikov[1978]}  for an equivalence Lie group it
is assumed $\xi^i_{\phi_k}=\zeta^j_{\phi_k}=0$. Discussion of the
generalization of the classical approach is given in \cite[Chapter~5,
Section~2.1]{bk:Meleshko[2005]}.}
 $x_i$, $u_j$
and $\phi _k$. Hereafter a sum over repeated indices is implied.

The coef\/f\/icients of the prolonged operator are obtained by
using the prolongation formulae:
\begin{gather*}
\zeta ^{u_{\alpha ,i}}=D_i^e\zeta ^{u_\alpha }-u_{\alpha
,j}D_i^e\xi
^{x_j} \qquad (i=1,2,3,4), \\
D_i^e=\partial _{x_i}+u_{\alpha ,i}\partial _{u_\alpha }+(\rho
_{x_i}W_{\beta ,1}+\dot \rho _{x_i}W_{\beta ,2})\partial _{W_\beta
},
\end{gather*}
where $\alpha =(\alpha _1,\alpha _2,\alpha _3,\alpha _4)$ and
$\beta =(\beta _1,\beta _2)$ are multiindices $(\alpha _i\geq 0$,
$\beta _i\geq 0)$,
\begin{gather*}
(\alpha _1,\alpha _2,\alpha _3,\alpha _4),\qquad j=(\alpha _1+\delta
_{1j},\alpha _2+\delta _{2j},\alpha _3+\delta _{3j},\alpha
_4+\delta _{4j}),
\\
 \ u_{(\alpha _1,\alpha _2,\alpha _3,\alpha _4)}= \frac{%
\partial ^{\alpha _1+\alpha _2+\alpha _3+\alpha _4}u}{\partial x_1^{\alpha
_1}\partial x_2^{\alpha _2}\partial x_3^{\alpha _3}\partial t^{\alpha _4}}%
,\qquad W_{(\beta _1,\beta _2)}=\frac{\partial ^{\beta _1+\beta _2}W}{%
\partial \rho ^{\beta _1}\partial \dot \rho ^{\beta _2}}.
\end{gather*}
The conditions that $W$ does not depend on $t$, $x_i$, $u_i$
($i=1,2,3$) give that
\begin{gather*}
\zeta _{x_i}^{u_k}=0,\qquad \zeta _{u_j}^{u_k}=0,\qquad \zeta
_{x_i}^W=0,\qquad \zeta _{u_j}^W=0 \qquad (i=1,2,3,4, \ j=1,2,3, \
k=4,5).
\end{gather*}
With these relations the prolongation formulae for the
coef\/f\/icients $\zeta ^{W_\beta }$ become:
\[
\zeta ^{W_{\beta ,i}}= \widetilde{D}_i^e\zeta ^{W_\beta }-W_{\beta ,1}%
\widetilde{D}_i^e\zeta ^{u_4}-W_{\beta ,2}\widetilde{D}_i^e\zeta
^{u_5} \qquad (i=1,2),
\]
where
\[
\widetilde{D}_1^e=\partial _\rho +W_{\beta ,1}\partial _{W_\beta
},\qquad \widetilde{D}_2^e=\partial _{\dot \rho }+W_{\beta
,2}\partial _{W_\beta }.
\]

For constructing the determining equations and solving them,
the symbolic computer  program Reduce~\cite{bk:Hearn} was applied.
Calculations yield the following basis of generators of the
equivalence Lie group
\begin{gather*}
X_1^e=\partial _{x_1},\qquad X_2^e=\partial _{x_2},\qquad
X_3^e=\partial _{x_3}, \qquad X_4^e=t\partial _{x_1}+\partial
_{u_1},\qquad X_5^e=t\partial _{x_2}+\partial
_{u_2},\\
X_6^e=t\partial _{x_3}+\partial _{u_3}, \qquad X_7^e=u_2\partial
_{u_2}-u_1\partial _{u_2}+x_2\partial _{x_1}-x_1\partial
_{x_2}, \\
X_8^e=u_3\partial _{u_1}-u_1\partial _{u_3}+x_3\partial
_{x_1}-x_1\partial _{x_3}, \qquad X_9^e=u_3\partial
_{u_2}-u_2\partial _{u_3}+x_3\partial _{x_2}-x_2\partial
_{x_3}, \\
X_{10}^e=\partial _t,\qquad X_{11}^e=t\partial _t+x_i\partial
_{x_i}, \qquad X_{12}^e=\partial _W,\qquad X_{13}^e=\rho \partial
_W,\qquad X_{14}^e=\dot \rho
\partial _W, \\
X_{15}^e=\dot \rho \partial _{\dot \rho }+\rho \partial _\rho
+W\partial _W, \qquad X_{16}^e=x_i\partial _{x_i}+u_i\partial
_{u_i}-2\rho \partial _\rho .
\end{gather*}
Here, only the essential part of the operators $X_i^e$ is written.
For example, the operator $X_{11}^e$ found as a result of the
calculations, is
\[
t\partial _t+x_i\partial _{x_i}-\dot \rho \partial _{\dot \rho }.
\]
The part $-\dot \rho \partial _{\dot \rho }$ is obtained from
$X_{11}^e$
using the prolongation formulae. 
The symmetry opera\-tors~$X^e_j$ $(1\leq j \leq 10)$ are symmetries
of the Galilean group\footnote{This group is admitted by many
systems of partial dif\/ferential equations applied in Newtonian
continuum mechanics. See, for example,
\cite{bk:Ovsiannikov[1978],bk:Handbook} and references therein.},
which are independent of a~potential function $W(\rho,\dot\rho)$.
The symmetries corresponding to the operators $X^e_1$, $X^e_2$,
$X^e_3$ are the space translation symmetries, $X^e_4$, $X^e_5$,
$X^e_6$ are the Galilean boosts, $X^e_7$, $X^e_8$ and $X^e_9$ are
the rotations and $X^e_{10}$ is the time translation symmetry. The
operator $X^e_{11}$ corresponds to a scaling symmet\-ry, which is
also admitted by the gas dynamics equations
\cite{bk:Ovsiannikov[1978]}. The symmetry corresponding to the
operator $X^e_{16}$ applies for a gas with a special state
equation~\cite{bk:Ovsiannikov[1978]}. Since the equivalence
transformations corresponding to the operators $ X_{11}^e$,
$X_{12}^e$, \dots, $X_{16}^e$ are applied for simplifying the
function $W$ in the process of the classif\/ication, let us
present these transformations. As the function $W$ depends on
$\rho$ and $\dot\rho$, only the transformations of these variables
are presented:
\begin{alignat*}{5}
& X_{11}^e: \quad && \rho ^{\prime }=\rho , \quad && \dot \rho ^{\prime }=e^{-a}\dot \rho ,\qquad && W^{\prime }=W; &\\
& X_{12}^e: && \rho ^{\prime }=\rho , && \dot \rho ^{\prime }=\dot
\rho , &&
W^{\prime }=W+a;& \\
& X_{13}^e: && \rho ^{\prime }=\rho , && \dot \rho ^{\prime }=\dot
\rho , &&
W^{\prime }=W+a\rho ;& \\
& X_{14}^e: && \rho ^{\prime }=\rho , && \dot \rho ^{\prime }=\dot
\rho , &&
W^{\prime }=W+a\dot \rho ; & \\
& X_{15}^e: && \rho ^{\prime }=e^a\rho , && \dot \rho ^{\prime
}=e^a\dot \rho ,
&& W^{\prime }=e^aW; & \\
& X_{16}^e: \quad && \rho ^{\prime }=e^{-2a}\rho , \qquad && \dot
\rho ^{\prime }=e^{-2a}\dot \rho ,\qquad && W^{\prime }=W.&
\end{alignat*}
Here $a$ is the group parameter.

\section{Admitted Lie group of (\ref{eq:main})}

An admitted generator $X$ of equations (\ref{eq:main}) is sought
in the form
\[
X=
\xi^{x_{1}}\partial_{x_{1}}+\xi^{x_{2}}\partial_{x_{2}}+\xi^{x_{3}}
\partial_{x_{3}}+ \xi^{t}\partial_{t}+\zeta^{u_{1}}\partial
_{u_{1}}+\zeta^{u_{2}}\partial _{u_{2}}+\zeta^{u_{3}}\partial
_{u_{3}}+\zeta^{\rho}\partial _{\rho},
\]
where the coef\/f\/icients of the generator are functions of the
variables $x_{1}$, $x_{2}$, $x_{3}$, $t$, $u_{1}$, $u_{2}$,
$u_{3}$, $\rho$.

Calculations showed that
\begin{gather*}
\xi ^{x_1}=c_6x_1t+c_4t+c_3x_3+x_1c_7+x_1c_1+c_5,
\\
\xi
^{x_2}=c_6x_2t+c_{12}t+x_3c_{11}+x_2c_7+x_2c_1-x_1c_{12}+c_{13},
\\
\xi ^{x_3}=c_6x_3t+c_{16}t+c_7x_3+c_1x_3-c_{11}x_2-c_3x_1+c_{17},
\\
\xi ^t=c_6t^2+c_7t+c_8,\qquad \zeta ^\rho =(-3c_6t+c_{15})\rho ,
\\
\zeta ^{u_1}=c_3u_3+c_2u_2-c_6u_1t+c_1u_1+c_6x_1+c_4,
\\
\zeta ^{u_2}c_{11}u_3-c_6u_2t+c_1u_2-c_2u_1+c_6x_2+c_{12},
\\
\zeta ^{u_3}=-c_6u_3t+c_1u_3-c_{11}u_2-c_3u_1+c_6x_3+c_{16},
\end{gather*}
where the constants $c_i$ $(i=1,2,\dots,8,11,12,13,15)$ satisfy
the conditions
\begin{gather}
27c_6\rho ^3(3W_{\dot \rho \rho \rho \rho }\dot \rho \rho +W_{\dot
\rho \rho \rho }\dot \rho -3W_{\rho \rho \rho }\rho -W_{\rho \rho
})+600W_{\dot \rho
\dot \rho }c_6\dot \rho ^2\rho  \nonumber\\
\qquad{}+25\dot \rho ^3(5W_{\dot \rho \dot \rho \dot \rho \dot
\rho }\dot \rho ^2(c_{15}-c_7)+5W_{\dot \rho \dot \rho \dot \rho
\rho }\dot \rho \rho
c_{15}+18W_{\dot \rho \dot \rho \rho }\rho c_{15} \nonumber\\
\qquad{}+W_{\dot \rho \dot \rho \dot \rho }\dot \rho
(28c_{15}-33c_7-10c_1)+18W_{\dot \rho \dot \rho
}(c_{15}-2c_7-2c_1))=0,\label{eq:fu(4)}
\\
W_{\dot \rho \dot \rho \dot \rho }\dot \rho
(c_7-c_{15})-c_{15}\rho W_{\dot \rho \dot \rho \rho
}+(2c_1-c_{15}+2c_7)W_{\dot \rho \dot \rho }+3c_6W_{\dot \rho \dot
\rho \dot \rho }\rho =0,\label{eq:fu(5)}
\\
9W_{\dot \rho \rho \rho \rho }\dot \rho \rho ^3c_{15}+40W_{\dot
\rho \dot \rho \dot \rho \dot \rho }\dot \rho
^4(c_7-c_{15})+W_{\dot \rho \dot \rho \dot \rho \rho }\dot \rho
^3\rho (9c_7-49c_{15})-9W_{\dot \rho \dot \rho
\rho \rho }\dot \rho ^2\rho ^2c_7 \nonumber\\
\qquad{}+8W_{\dot \rho \dot \rho \dot \rho }\dot \rho
^3(10c_1-17c_{15}+22c_7)+2W_{\dot \rho \dot \rho \rho }\dot \rho
^2\rho
(9c_1-37c_{15}+9c_7)-9W_{\rho \rho \rho }\rho ^3c_{15} \nonumber\\
\qquad{}+9W_{\dot \rho \rho \rho }\dot \rho \rho
^2(c_{15}-2c_1)+56W_{\dot \rho \dot \rho }\dot \rho
^2(2c_1-c_{15}+2c_7)+9W_{\rho \rho }\rho
^2(2c_1-c_{15})=0,\label{eq:fu(6)}
\\
c_6(5W_{\dot \rho \dot \rho \dot \rho }\dot \rho +3W_{\dot \rho
\dot \rho \rho }\rho +5W_{\dot \rho \dot \rho
})=0.\label{eq:fu(8)}
\end{gather}

The determining equations (\ref{eq:fu(4)})--(\ref{eq:fu(8)})
def\/ine the kernel of admitted Lie algebras and its extensions.
The kernel of admitted Lie algebras consists of the generators
\begin{gather*}
Y_1=\partial _{x_1},\qquad Y_2=\partial _{x_2},\qquad Y_3=\partial
_{x_3},\qquad
Y_{10}=\partial _t, \\
Y_4=t\partial _{x_1}+\partial _{u_1},\qquad Y_5=t\partial
_{x_2}+\partial
_{u_2},\qquad Y_6=t\partial _{x_3}+\partial _{u_3}, \\
Y_7=x_2\partial _{x_3}-x_3\partial _{x_2}+u_2\partial
_{u_3}-u_3\partial
_{u_2}, \\
Y_8=x_3\partial _{x_1}-x_1\partial _{x_3}+u_3\partial
_{u_1}-u_1\partial
_{u_3}, \\
Y_9=x_1\partial _{x_2}-x_2\partial _{x_1}+u_1\partial
_{u_2}-u_2\partial _{u_1}.
\end{gather*}
Extensions of the kernel depend on the value of the function
$W(\rho ,\dot \rho )$. They can only be operators of the form
\[
c_1X_1+c_6X_6+c_7X_7+c_{15}X_{14},
\]
where
\begin{gather*}
X_1=x_i\partial _{x_i}+u_i\partial _{u_i}, \qquad X_6=t(t\partial
_t+x_i\partial _{x_i}-u_i\partial _{u_i}-3\rho \partial
_\rho )+x_i\partial _{u_i} \\
X_7=x_i\partial _{x_i}+t\partial t,\qquad X_9=x_2\partial
_{x_2}+u_2\partial _{u_2},\qquad X_{14}=\rho \partial _\rho .
\end{gather*}
Relations between the constants $c_1$, $c_6$, $c_7$, $c_{15}$
depend on the function $W(\rho ,\dot \rho )$.

\subsection[Case $c_{6}\ne 0$]{Case $\boldsymbol{c_{6}\ne 0}$}

Let $c_6\ne 0$, then equation (\ref{eq:fu(8)}) gives
\[
5W_{\dot \rho \dot \rho \dot \rho }\dot \rho +3W_{\dot \rho \dot
\rho \rho }\rho +5W_{\dot \rho \dot \rho }=0.
\]
The general solution of this equation is 
$W_{\dot \rho \dot \rho }=\rho ^{-5/3}g(\dot \rho \rho ^{-5/3}),$
where the function $g$ is an arbitrary function of integration.
Substitution of $W_{\dot \rho \dot \rho }$ into
equation~(\ref{eq:fu(5)}) shows that the function $g=2q_0$ is
constant. Hence,
\[
W=q_0\dot {\rho ^2}\rho ^{-5/3}+\varphi _1(\rho )\dot \rho
+\varphi _2(\rho ),
\]
where the functions $\varphi _2(\rho )$ and $\varphi _1(\rho )$
are arbitrary. Substituting this potential function in the other
equations (\ref {eq:fu(4)})--(\ref{eq:fu(6)}), one obtains
\[
3\rho \varphi _2^{\prime \prime \prime }+\varphi _2^{\prime \prime
}=0,\qquad (c_7+2c_1)\varphi _2^{\prime \prime }=0,\qquad
c_{15}=-3(c_{1}+c_{7}).
\]

If $\varphi _2^{\prime \prime }=0$, then the extension of the
kernel of admitted Lie algebras is given by the generators
\[
X_6,\qquad X_1-3X_{14},\qquad X_7-3X_{14}.
\]

If $\varphi _2^{\prime \prime }=C_2\rho ^{-3}\neq 0$, then the
extension of the kernel is given by the generators
\[
X_6,\qquad X_1-2X_7+3X_{14}.
\]

\subsection[Case $c_{6}=0$]{Case $\boldsymbol{c_{6}=0}$}

Let $c_6=0$, then equation (\ref{eq:fu(5)}) becomes
\begin{gather}
\label{eq:fu(5)a} -c_{15}a+(c_1+c_7)b+c_7c=0,
\end{gather}
where
\[
a= \dot \rho W_{\dot \rho \dot \rho \dot \rho }+\rho W_{\dot \rho
\dot \rho \rho }+W_{\dot \rho \dot \rho },\qquad b=2W_{\dot \rho
\dot \rho },\qquad c=\dot \rho W_{\dot \rho \dot \rho \dot \rho }.
\]
Further analysis of the determining equations
(\ref{eq:fu(4)})--(\ref{eq:fu(6)}) is similar to the group
classif\/ication of the gas dynamics
equations~\cite{bk:Ovsiannikov[1978]}.

Let us analyze the vector space ${\rm Span}(V)$, where the set $V$
consists of vectors $(a,b,c)$ with $\rho $ and $\dot \rho $ are
changed. If the function $W(\rho ,\dot \rho )$ is such that $\dim
({\rm Span}(V))=3$, then equation (\ref{eq:fu(5)a}) is only
satisf\/ied for
\[
c_1=0,\qquad c_7=0,\qquad c_{15}=0,
\]
which does not give extensions of the kernel of admitted Lie
algebras. Hence, one needs to study $\dim ({\rm Span}(V))\leq 2$.

\subsubsection[Case $\dim ({\rm Span}(V))=2$]{Case $\boldsymbol{\dim ({\rm Span}(V))=2}$}

Let $\dim ({\rm Span}(V))=2$. There exists a constant vector
$(\alpha ,\beta ,\gamma )\neq 0,$ which is orthogonal to the set
$V$:
\begin{gather}
\label{eq:nov30.1}\alpha a+\beta b+\gamma c=0.
\end{gather}
This means that the function $W(\rho ,\dot \rho )$ satisf\/ies the
equation
\begin{gather}
\label{eq:k2eq0.2}(\alpha +\gamma )\dot \rho W_{\dot \rho \dot
\rho \dot \rho }+\alpha \rho W_{\rho \dot \rho \dot \rho
}=-(\alpha +2\beta )W_{\dot \rho \dot \rho }.
\end{gather}
The characteristic system of this equation is
\[
\frac{d\dot \rho }{(\alpha +\gamma )\dot \rho }=\frac{d\rho
}{\alpha \rho }= \frac{dW_{\dot \rho \dot \rho }}{-(\alpha +2\beta
)W_{\dot \rho \dot \rho }} .
\]
The general solution of equation (\ref{eq:k2eq0.2}) depends on the
values of the constants $\alpha$, $\beta $ and $\gamma $.

\paragraph{Case $\boldsymbol{\alpha =0}$.}
Because of equation (\ref{eq:nov30.1}) and the condition $W_{\dot
\rho \dot \rho }\neq 0$, one has $\gamma \neq 0$. The general
solution of equation (\ref{eq:k2eq0.2}) is
\begin{gather}
\label{eq:k2eq0.3}W_{\dot \rho \dot \rho }(\rho ,\dot \rho
)=\tilde \varphi \dot \rho ^k,
\end{gather}
where $k=-2\beta /\gamma $, and $\widetilde{\varphi }$ is an
arbitrary function of integration. Substitution of
(\ref{eq:k2eq0.3}) into (\ref {eq:fu(5)a}) leads to
\begin{gather}
\label{eq:k2eq0.4}c_{15}\rho \tilde \varphi ^{\prime }-\tilde
\varphi (\rho )\left( 2c_1-(k+1)c_{15}+(k+2)c_7\right) =0.
\end{gather}
If $c_{15}\neq 0$, the dimension $\dim ({\rm Span}(V))=1,$ which
contradicts to the
assumption. Hence, $c_{15}=0$ and from (\ref{eq:k2eq0.4}) one obtains $%
\tilde {c_1}=-(k+2)c_7/2$. The extension of the kernel in this
case is given by the generator
\[
-pX_1+2X_7,
\]
where $p=k+2$.

If $(k+2)(k+1)\neq 0$, then integrating (\ref{eq:k2eq0.3}), one
f\/inds
\begin{gather*}
W(\rho ,\dot \rho )=\varphi (\rho )\dot \rho ^p+\varphi _1(\rho
)\dot \rho +\varphi _2(\rho ),
\end{gather*}
where $\varphi _1(\rho )$ and $\varphi _2(\rho )$ are arbitrary
functions. Substituting this function $W$ into
(\ref{eq:fu(4)})--(\ref{eq:fu(6)}) one has $\varphi _2^{\prime
\prime }=0$.

If $k=-2$, then
\begin{gather*}
W(\rho ,\dot \rho )=\varphi (\rho )\ln (\dot \rho )+\dot\rho
\varphi _1(\rho ) +\varphi _2(\rho ),
\end{gather*}
and  $\varphi _2^{\prime \prime }=0$, similar to the previous
case.

If $k=-1$, then
\begin{gather*}
W(\rho ,\dot \rho )=\varphi (\rho )\dot \rho \ln (\dot \rho )+\dot
\rho \varphi _1(\rho )+\varphi _2(\rho ),
\end{gather*}
and also $\varphi _2^{\prime \prime }=0$.

\paragraph{Case $\boldsymbol{\alpha \neq 0}$.}

The general solution of equation (\ref{eq:k2eq0.3}) is
\begin{gather}
\label{eq:case2} W_{\dot \rho \dot \rho }(\rho ,\dot \rho
)=\varphi (\dot \rho \rho ^k)\rho ^\lambda ,
\end{gather}
where $k=-(1+\gamma /\alpha )$, $\lambda =-(1+2\beta /\alpha )$
and $\varphi $ is an arbitrary function. Substitution of this
function into (\ref {eq:fu(5)a}) leads to
\begin{gather*}
k_0\varphi ^{\prime }z+k_1\varphi =0,
\end{gather*}
where
\[
z=\dot \rho \rho ^k,\qquad k_0=c_7-c_{15}(k+1),\qquad
k_1=2c_1-c_{15}(\lambda +1)+2c_7.
\]
Since $\dim ({\rm Span}(V))=2$, one obtains that $k_0=0$ and
$k_1=0$ or
\[
c_7=c_{15}(k+1),\qquad c_1=c_{15}(p-1)/2,
\]
where $p=\lambda -2k$. Integrating (\ref{eq:case2}), one f\/inds
\begin{gather}
\label{eq:case2n}W(\rho ,\dot \rho )=\rho ^p\varphi (\dot \rho
\rho ^k)+\dot \rho \varphi _1(\rho )+\varphi _2(\rho ).
\end{gather}
Substitution of (\ref{eq:case2n}) into
(\ref{eq:fu(4)})--(\ref{eq:fu(6)}) gives
\begin{gather*}
\rho \varphi _2^{\prime \prime \prime }+(2k-\lambda +2)\varphi
_2^{\prime \prime }=0.
\end{gather*}
Solving this equation, one has
\[
\varphi _2^{\prime \prime }=C_2\rho ^{p-2},
\]
where $C_2$ is an arbitrary constant. The extension of the kernel
is given by the generator
\[
(p-1)X_1+2(k+1)X_7+2X_{14}.
\]

\subsubsection[Case $\dim ({\rm Span}(V))=1$]{Case $\boldsymbol{\dim ({\rm Span}(V))=1}$}

Let $\dim ({\rm Span}(V))=1$. There exists a constant vector
$(\alpha ,\beta ,k)\neq 0$ such that
\[
(a,b,c)=(\alpha ,\beta ,k)B
\]
with some function $B(\rho ,\dot \rho )\neq 0$. Because $W_{\dot
\rho \dot \rho }\neq 0$, one has that $\beta \neq 0$. Hence, the
function $W(\rho ,\dot \rho )$ satisf\/ies the equations
\[
\dot \rho W_{\dot \rho \dot \rho \dot \rho }+\rho W_{\rho \dot
\rho \dot \rho }+(1-2\tilde \alpha )W_{\dot \rho \dot \rho
}=0,\qquad \dot \rho W_{\dot \rho \dot \rho \dot \rho }-2\gamma
W_{\dot \rho \dot \rho }=0.
\]

The general solution of the latter equation is
\[
W_{\dot \rho \dot \rho }(\rho ,\dot \rho )=\varphi (\rho )\dot
\rho ^k
\]
with arbitrary function $\varphi (\rho )$. Substituting this
solution into the f\/irst equation, one obtains
\[
\rho \varphi ^{\prime }(\rho )+(1-2\tilde \alpha +k)\varphi (\rho
)=0, \qquad
\tilde{\alpha}=\alpha/\beta.
\]
Thus,
\begin{gather}
\label{eq:case3} W_{\dot \rho \dot \rho }=-q_0\dot \rho ^k\rho
^\lambda ,
\end{gather}
where $\lambda =-(1-2\tilde \alpha +k)$, $q_0$ is an arbitrary
constant. Since $\dim ({\rm Span}(V))=1$, then $q_0\neq 0$,
$\lambda $~and~$k$ are such that $\lambda ^2+k^2\neq 0$.

Substituting (\ref{eq:case3}) into (\ref{eq:fu(5)a}), it becomes
\begin{gather*}
-c_{15}(k+\lambda +1)+c_7(k+2)+2c_1=0.
\end{gather*}
Integration of (\ref{eq:case3}) depends on the quantity of $k$.

If $(k+2)(k+1)\neq 0$, then integrating (\ref{eq:case3}), one
obtains
\begin{gather*}
W(\rho ,\dot \rho )=-q_0\rho ^\lambda \dot \rho ^p+\dot \rho
\varphi _1(\rho )+\varphi _2(\rho ),\qquad p(p-1)\neq 0,
\end{gather*}
where $p=k+2$. Substituting this $W$ into equations
(\ref{eq:fu(4)})--(\ref {eq:fu(6)}), one obtains
\[
c_1=\left( c_{15}(p+\lambda -1)-c_7p)\right) /2,
\]
with the function $\varphi _2(\rho )$ satisfying the condition
\[
c_{15}\rho \varphi _2^{\prime \prime \prime }+\varphi _2^{\prime
\prime }(-c_{15}(p+\lambda -2)+c_7p)=0.
\]

If $\varphi _2^{\prime \prime }=C_2\rho ^{-\mu }\neq 0$, the
extension of the kernel is given by the generator
\[
(1-\mu )X_1+2(X_{14}+\phi X_7),
\]
where $\phi =(\mu +\lambda +p-2)/p$. If $\varphi _2^{\prime \prime
}=0$, the extension is given by the generators
\[
pX_1-2X_7,\qquad (p+\lambda -1)X_1+2X_{14}.
\]

If $k=-2$, then integrating (\ref{eq:case3}), one obtains
\begin{gather*}
W(\rho ,\dot \rho )=-q_0\rho ^\lambda \ln (\dot \rho )+\dot \rho
\varphi _1(\rho )+\varphi _2(\rho ),\qquad q_0\neq 0.
\end{gather*}
Substituting this into equations
(\ref{eq:fu(4)})--(\ref{eq:fu(6)}), we obtain
\[
c_1=c_{15}(\lambda -1)/2,
\]
and the condition
\[
c_{15}(\rho \varphi _2^{\prime \prime \prime }-\varphi _2^{\prime
\prime }(\lambda +2))+q_0\lambda (\lambda -1)(c_{15}-c_7)\rho
^{\lambda -2}=0.
\]

If $\lambda (\lambda -1)=0$ and $\varphi_{2}$ is arbitrary, then
the extension is given only by the generator
\[
X_7.
\]

If $\lambda (\lambda -1)=0$ and $\varphi _2^{\prime \prime
}=C_2\rho ^{\lambda +2}$, then the extension of the kernel
consists of the generators
\[
(\lambda -1)X_1+2X_{14},\qquad X_7.
\]

If $\lambda (\lambda -1)\neq 0$ and $\varphi _2^{\prime \prime
}=C_2\rho ^{\lambda +2}-\frac{q_0}4\lambda (\lambda -1)\mu \rho
^{\lambda -2}$, then the extension is
\[
(\lambda -1)X_1+2(X_{14}+(\mu +1)X_7),
\]
where $c_7=(\mu +1)c_{15}$.

If $k=-1$, then integrating (\ref{eq:case3}), one obtains
\begin{gather*}
W(\rho ,\dot \rho )=-q_0\rho ^\lambda \dot \rho \ln (\dot \rho
)+\dot \rho \varphi _1(\rho )+\varphi _2(\rho ),
\end{gather*}
and substituting it into equations
(\ref{eq:fu(4)})--(\ref{eq:fu(6)}), we obtain
\[
c_1=(c_{15}\lambda -c_7)/2,
\]
and the condition
\[
c_{15}\rho \varphi _2^{\prime \prime \prime }+\varphi _2^{\prime
\prime }(-c_{15}\lambda +c_{15}+c_7)=0.
\]
One needs to study two cases. If $\varphi _2^{\prime \prime }\neq
0$, then the extension is possible only for $\varphi _2=C_2\rho
^{-\mu }\neq 0$, where $\mu =-\lambda +1+c_7/c_{15}$. The
extension of the kernel is given by the generator
\[
(1-\mu )X_1+2(\mu +\lambda -1)X_7+2X_{14}.
\]
If $\varphi _2^{\prime \prime }=0$, then the extension of the
kernel consists of the generators
\[
X_1-2X_7,\qquad X_{14}+\lambda X_7.%
\]

\subsubsection[Case $\dim ({\rm Span}(V))=0$]{Case $\boldsymbol{\dim ({\rm Span}(V))=0}$}

Let $\dim ({\rm Span}(V))=0$. The vector $(a,b,c)$ is constant:
\[
(a,b,c)=(\alpha ,\beta ,k)
\]
with some constant values $\alpha$, $\beta $ and $k$. This leads to
\[
W_{\dot \rho \dot \rho }=-2q_0,
\]
where $q_0\neq 0$ is constant. Integrating this equation, one
obtains
\begin{gather}
\label{eq:case4}W(\rho ,\dot \rho )=-q_0\dot \rho ^2+\dot \rho
\varphi _1(\rho )+\varphi _2(\rho ).
\end{gather}
Substituting (\ref{eq:case4}) into equation
(\ref{eq:fu(4)})--(\ref{eq:fu(6)}), we obtain
\[
c_1=(c_{15}-2c_7)/2,
\]
and the condition
\[
c_{15}\rho \varphi _2^{\prime \prime \prime }+2c_7\varphi
_2^{\prime \prime }=0.
\]

If $\varphi _2^{\prime \prime }\neq 0$, then $\varphi _2=C_2\rho
^{-\mu }$, where $\mu =2c_7/c_{15}$. The extension of the kernel
consists of the generator
\[
(1-\mu )X_1+2X_{14}+\mu X_7.
\]

If $\varphi _2^{\prime \prime }=0$, then the extension of the
kernel is given by the generators
\[
X_1+2X_{14},\qquad X_1-X_7.
\]
The result of group classif\/ication of equations (\ref{eq:main})
is summarized in Table \ref{tab:grclass}. The linear part with
respect to $\dot \rho $ of the function $W(\rho ,\dot \rho )$ is
omitted. Notice also that the change $t\rightarrow -t$ has to
conserve the potential function $W$, this leads to $\varphi
_1(\rho )=0$.


\begin{remark} The Green--Naghdi model belongs to the class $M_7$ in Table~\ref{tab:grclass} with $\lambda = 1$, $p=2$ and $\mu =0$.
Invariant solutions of the one-dimensional Green--Naghdi model
completely studied in~\cite{bk:BagderinaChupakhin}.
\end{remark}

\begin{remark} The one-velocity dissipation-free Iordanski--Kogarko--Wijngaarden model has an extension of the kernel of
admitted Lie algebras only for a special internal energy of the
gas phase (class $M_3\;(p=2)$ in Table \ref{tab:grclass}), which
corresponds to a Chaplygin gas $\varepsilon _{20}\left( \rho
_{20}\right) =\gamma _1/\rho _{20}+\gamma _0$, where $\gamma _1$
and $\gamma _0$ are constants.
\end{remark}

\begin{table}[t]\centering
\caption{Group classif\/ication of equations (\ref{eq:main}).}
\vspace{1mm}

 \label{tab:grclass}
\begin{tabular}{|r|l|l|l|}
\hline \tsep{0.5ex}\bsep{0.5ex} & $W(\rho ,\dot \rho )$ &
Extensions & Remarks
\\
\hline \tsep{0.5ex}\bsep{0.5ex} $M_1$& $-q_0
\rho^{-5/3}\dot{\rho}^{2}+\varphi_{2}(\rho)$&$ X_6$, $X_1-2X_7
+3X_{14} $&  $\varphi _2^{\prime \prime }=C_2\rho^{-3}\neq 0$
\\
\hline \tsep{0.5ex}\bsep{0.5ex} $M_2$ & $ $ & $ X_6$,
$X_1-3X_{14}$, $X_{7}-3X_{14}$ & $ \varphi_{2}''= 0 $
\\
\hline \tsep{0.5ex}\bsep{0.5ex} $M_3$&
$\varphi(\rho)\dot{\rho}^{p}+\varphi_{2}$ & $ -pX_1+2X_7 $ & $
\varphi_{2}''=0$
\\
\hline \tsep{0.5ex}\bsep{0.5ex} $M_4$ &
$\varphi(\rho)\ln\dot{\rho}+\varphi_{2}$ & $ X_7$ &$\varphi_2''=0$
\\
\hline \tsep{0.5ex}\bsep{0.5ex} $M_5$ &
$\dot{\rho}\varphi(\rho)\ln \dot{\rho}+\varphi_{2} $ & $X_1-2X_7$
&$\varphi_2''=0$
\\
\hline \tsep{0.5ex}\bsep{0.5ex} $M_6$ &
$\rho^{p}\varphi(\dot{\rho}\rho^{k})+\varphi_{2}$ &
$(p-1)X_1+2(X_7(k+1)+X_{14})$ & $ \varphi_{2}''=C_2\rho^{p-2}$
\\ \hline
\tsep{0.5ex} $M_7$ & $
-q_0\rho^{\lambda}\dot{\rho}^{p}+\varphi_{2}$ &
$(1-\mu)X_1+2(X_{14}+\phi X_7)$
&$\varphi _2^{\prime \prime }=C_2\rho ^{-\mu }\neq 0$, \\
&&&$p(p-1)\neq 0$,\\
&&&$\phi =(\mu+\lambda +p-2)/p$\bsep{0.5ex}\\
 \hline
\tsep{0.5ex} $M_8$
&   $ $ & $ pX_1-2X_7$, &$\varphi_{2}''=0$, \\
&&$(p+\lambda -1)X_1+2X_{14}$&$p(p-1)\neq 0$\bsep{0.5ex}
\\
\hline \tsep{0.5ex} $M_9$ &
$-q_0\rho^{\lambda}\ln\dot{\rho}+\varphi_{2} $ & $X_7$ &
$\varphi_{2}(\rho)$ arbitrary,
\\
 & & & $\lambda(\lambda-1)=0$\bsep{0.5ex}
\\
\hline \tsep{0.5ex} $M_{10}$ & & $ (\lambda-1) X_1+2X_{14}$, &
$\varphi_2^{\prime \prime}=C_2\rho^{\lambda+2}$,
\\
 &&$X_{7}$& $\lambda(\lambda-1)=0$\bsep{0.5ex}
\\
\hline \tsep{0.5ex} $M_{11}$ & & $ (\lambda-1) X_1+2(X_{14}+(\mu
+1)X_7) $& $\varphi_2^{\prime \prime}= C_2\rho^{\lambda+2}$
\\&&
 &$-\frac{q_0}{4}\lambda(\lambda-1)\mu\rho^{\lambda-2}$,\\
&&&$\lambda(\lambda-1)\neq 0$\bsep{0.5ex}
\\
\hline \tsep{0.5ex} $M_{12}$ &
$-q_0\rho^{\lambda}\dot{\rho}\ln\dot{\rho}+\varphi_{2} $ & $(1-\mu
)X_1+2(\mu +\lambda -1)X_7+2X_{14}$ & $\varphi_2 = C_2\rho^{-\mu}\neq
0$\bsep{0.5ex}
\\
 \hline
\tsep{0.5ex} $M_{13}$ &  &$X_1-2X_7$,\ $X_{14}+\lambda X_{7}$ &
$\varphi_{2}''= 0$\bsep{0.5ex}
\\
\hline \tsep{0.5ex} $M_{14}$ & $-q_0\dot{\rho^{2}}+\varphi_{2} $ &
$ (1-\mu )X_1+2X_{14}+\mu X_7$ & $\varphi_2=C_2\rho^{-\mu}\neq
0$\bsep{0.5ex}
\\
\hline \tsep{0.5ex} $M_{15}$ &   $ $ & $ X_1+2X_{14}, $ \
$X_1-X_7$& $ \varphi _2''=0$\bsep{0.5ex}
\\
\hline
\end{tabular}
\end{table}

\section{Special vortex}

In this section a special vortex solution is considered. With the
spherical coordinates \cite{bk:Ovsiannikov[1995]}$:$
\begin{gather*}
x=r\sin \theta \cos \varphi ,\qquad y=r\sin \theta \sin \varphi
,\qquad z=r\cos \theta ,
\\
U=u\sin \theta \cos \varphi +v\sin \theta \sin \varphi +w\cos \theta , \\
U_2=u\cos \theta \cos \varphi +v\cos \theta \sin \varphi -w\sin \theta , \\
U_3=-u\sin \varphi +v\cos \varphi ,
\end{gather*}
the generators $X_7$, $X_8$, $X_9$ are
\begin{gather*}
X_7=-\sin \varphi \partial _\theta -\cos \varphi \cot \theta
\partial _\varphi +\cos \varphi (\sin \theta )^{-1}(U_2\partial
_{U_3}-U_3\partial
_{U_2}), \\
X_8=-\cos \varphi \partial _\theta -\sin \varphi \cot \theta
\partial _\varphi +\sin \varphi (\sin \theta )^{-1}(U_2\partial
_{U_3}-U_3\partial _{U_2}), \qquad X_9=\partial _\varphi .
\end{gather*}
Introducing cylindrical coordinates $(H,\omega )$ into the
two-dimensional space of vectors $(U_2,U_3)$
\[
U_2=H\cos \omega ,\qquad U_3=H\sin \omega ,
\]
the f\/irst two generators become
\begin{gather*}
X_7=-\sin \varphi \partial _\theta -\cos \varphi \cot \theta
\partial
_\varphi +\cos \varphi (\sin \theta )^{-1}\partial _\omega , \\
X_8=-\cos \varphi \partial _\theta -\sin \varphi \cot \theta
\partial _\varphi +\sin \varphi (\sin \theta )^{-1}\partial
_\omega .
\end{gather*}
The singular vortex solution \cite{bk:Ovsiannikov[1995]} is
def\/ined by the representation
\begin{gather*}
U=U(t,r),\qquad H=H(t,r),\qquad \rho =\rho (t,r),\qquad \omega
=\omega (t,r,\theta ,\varphi ).
\end{gather*}
The function $\omega (t,r,\theta ,\varphi )$ is ``superf\/luous'':
it depends on all independent variables. If $H=0$, then the
tangent component of the velocity vector is equal to zero. This
corresponds to the spherically symmetric f\/lows. For a singular
vortex, it is assumed that $H\neq 0.$

In a manner similar to \cite{bk:Ovsiannikov[1995]} one f\/inds
that for system (\ref{eq:main}), the invariant functions $U(t,r)$,
$H(t,r)$\ and $\rho (t,r)$ have to satisfy the system of partial
dif\/ferential equations with the two independent variables $t$
and $r$:
\begin{gather}
r^2D_0\rho +\rho (r^2U)_r=\rho \alpha h,\qquad D_0U+\rho
^{-1}p_r=r^{-3}\alpha
^2, \nonumber\\
D_0h=r^{-2}\alpha (h^2+1),\qquad D_0\alpha =0, \nonumber\\
p=\rho (W_\rho -\dot \rho W_{\rho \dot \rho }-W_{\dot \rho \dot
\rho }D_0\dot \rho )+W_{\dot \rho }\dot \rho -W,\label{eq:spvort}
\end{gather}
where $\alpha =rH$, $D_0=\partial _t+U\partial _r$, and the
function $h(t,r)$ is introduced for convenience during the
compatibility analysis.

The equivalence Lie group of equations (\ref{eq:spvort})
corresponds to the generators
\begin{gather*}
X_0^e=\partial _t,\qquad X_2^e=\rho \partial _W,\qquad
X_3^e=2t\partial _t-U\partial _U-3\rho \partial _\rho -5 {\dot
\rho }\partial _{\dot \rho }-3W\partial _W, \\ X_4^e={\dot \rho }
\partial _{\dot \rho }+\rho \partial _\rho +W\partial _W,\qquad X_5^e=x\partial
_x+U\partial _U+2\alpha \partial _\alpha +2W\partial _W.
\end{gather*}

Calculations yield that the kernel of admitted Lie algebras
consists of the generator
\[
X_0=\partial _t,
\]
extensions of the kernel can only be operators of the form
\[
k_1X_1+k_2X_2+k_3X_3+k_4X_4,
\]
where
\begin{gather*}
X_1=t\partial _t-U\partial _U-\alpha \partial _\alpha + {\dot \rho
}\partial _{\dot \rho },\qquad X_2=t(t\partial _t+r\partial
_r-U\partial _U-3\rho \partial _\rho -5{\dot \rho }\partial _{\dot
\rho })+r\partial _U-3\rho \partial _{\dot \rho }, \\
X_3=2t\partial _t+r\partial _r-U\partial _U-3\rho \partial _\rho
-5{\dot \rho }\partial _{\dot \rho },\qquad X_4={\dot \rho
}\partial _{\dot \rho }+\rho \partial _\rho .
\end{gather*}
The constants $k_i$ ($i=1,2,3,4$) depend on the function $W(\rho
,{\dot \rho })$. These extensions are presented in
Table~\ref{tab:grclvort}.

\begin{table}[t]\centering
\caption{Group classif\/ication of equations (\ref{eq:spvort}).}
 \label{tab:grclvort}
 \vspace{1mm}

\begin{tabular}{|r|l|l|l|}
\hline \tsep{0.5ex} \bsep{0.5ex} &$ W(\rho ,\dot \rho ) $&
Extensions & Remarks
\\ \hline
\tsep{0.5ex} \bsep{0.5ex} $M_1 $&$ -q_0\dot \rho ^2\rho
^{-5/3}+\beta \rho ^{5/3} $&$ X_2$, $X_3 $& $ q_0\beta \neq 0 $
\\ \hline
\tsep{0.5ex} \bsep{0.5ex} $M_2 $&$ -q_0 \dot \rho ^2\rho ^{-5/3}
$&$ X_2$, $X_1$, $X_3 $&$ q_0 \neq 0$
\\ \hline
\tsep{0.5ex} \bsep{0.5ex} $M_3 $&$ \varphi (\rho )\dot \rho ^p $&$
X_1+(2-p)X_3 $&$ p(p-1)\neq 0 $
\\ \hline
\tsep{0.5ex} \bsep{0.5ex} $M_4 $&$ -(q_0 \rho +\gamma )\ln (\dot
\rho )+\varphi _2(\rho ) $&$ X_1+X_3 $&$ \varphi _2$ arbitrary
\\ \hline
\tsep{0.5ex} \bsep{0.5ex} $M_5 $&$ \varphi (\rho )\dot \rho \ln
(\dot \rho ) $&$ 2X_1+X_3 $&$  $
\\ \hline
\tsep{0.5ex} \bsep{0.5ex} $M_6 $&$ \rho ^\lambda \varphi (\dot
\rho \rho ^k)+\varphi _2(\rho ) $&$ 2X_1-(\lambda
-2)X_3,\;X_4-kX_3 $&$ \varphi _2^{\prime \prime }=C_2\rho
^{\lambda -2} $
\\ \hline
\tsep{0.5ex}  $M_7 $&$ -q_0 \rho ^\lambda \dot \rho ^p+\varphi
_2(\rho ) $&$ 2(\mu X_1+2(2\mu +p(\lambda -\mu ))X_3 $&$ \varphi
_2^{\prime \prime }=C_2\rho ^\mu$
\\
& &$ +(2-\lambda )(2X_1+(2-p)X_3) $&$ p(p-1)\neq 0 $\bsep{0.5ex}
\\ \hline
\tsep{0.5ex}  $M_8 $&$ -q_0 \rho ^\lambda \dot \rho ^p $&$
-2X_1+(2-\lambda )X_3, $&$ p(p-1)\neq 0 $
\\
& &$ (p-2)X_3-2X_6 $&\bsep{0.5ex}
\\ \hline
\tsep{0.5ex}  $M_9 $&$ -q_0 \rho ^\lambda \ln (\dot \rho )+\varphi
_2(\rho ) $&$ X_1+X_3 $&$ \varphi _2$ arbitrary
\\
&  &  &$ \lambda (\lambda -1)=0$\bsep{0.5ex}
\\ \hline
\tsep{0.5ex} $M_{10} $& &$ X_3+X_6$, $2X_1+(\lambda -1)X_3 $&$
\varphi _2^{\prime \prime }=C_2\rho ^{\lambda -2} $
\\ \hline
&  &  &$ \lambda (\lambda -1)=0 $ \bsep{0.5ex}
\\ \hline
\tsep{0.5ex}  $M_{11} $&  &$ X_1+\frac{\lambda -1}2X_3+X_6$&$
\varphi _2^{\prime \prime }=\rho ^{\lambda -2}(q_0 \ln (\rho
)+\beta ) $
\\
&  & $+\frac \alpha {C\lambda (\lambda -1)}\left( X_3+X_6\right) $
&$ \lambda (\lambda -1)\neq 0 $\bsep{0.5ex}
\\ \hline
\tsep{0.5ex}  $M_{12} $&$ -q_0 \rho ^\lambda \dot \rho \ln (\dot
\rho )+\varphi _2(\rho ) $&$ 2X_1+\lambda X_3 $&$ \varphi
_2^{\prime \prime
}=C_2\rho ^\mu \neq 0$\\
&&$+(\lambda -\mu -1)(X_3+2X_6)$&  \bsep{0.5ex}
\\ \hline
\tsep{0.5ex} \bsep{0.5ex} $M_{13} $&$ -q_0 \rho ^\lambda \dot \rho
\ln (\dot \rho ) $&$ 2X_1+X_3$, $X_4 $&
\\ \hline
\tsep{0.5ex} \bsep{0.5ex} $M_{14} $&$ -q_0 \dot \rho ^2+\varphi
_2(\rho ) $&$ 2X_1+X_3-\mu X_6 $&$ \varphi _2^{\prime \prime
}=C_2\rho ^\mu \neq 0 $
\\ \hline
\tsep{0.5ex} \bsep{0.5ex} $M_{15} $&$ -q_0 \dot \rho ^2 $&$ X_1$,
$X_4 $&
\\ \hline
\end{tabular}
\end{table}


\subsection{Steady-state special vortex}

Let us consider the invariant solution corresponding to the kernel
$\{X_0\}$. This type of solution for the gas dynamics equations
was studied in \cite{bk:Chupakhin[2003]}. The representation of
the solution is
\[
\rho =\rho (r),\qquad U=U(r),\qquad h=h(r),\qquad \alpha =\alpha
(r).
\]
Equations (\ref{eq:spvort}) become
\begin{gather}
U\rho ^{\prime }+\rho (r^2U)^{\prime }=\rho \alpha h,\qquad
UU^{\prime }+\rho
^{-1}p^{\prime }=r^{-3}\alpha ^2,\nonumber\\
Uh^{\prime }=r^{-2}\alpha (h^2+1),\qquad U\alpha ^{\prime }=0, \nonumber\\
p=\rho (W_\rho -U\rho ^{\prime }W_{\rho \dot \rho }-W_{\dot \rho
\dot \rho }U(U\rho ^{\prime })^{\prime })+W_{\dot \rho }U\rho
^{\prime }-W, \qquad \dot \rho =U\rho'.\label{eq:spvortst}
\end{gather}
In~\cite{bk:Chupakhin[2003]} it is shown that for the gas dynamics
equations all dependent variables can be represented through the
function~$h(r)$, which satisf\/ies a f\/irst-order ordinary
dif\/ferential equation. Here also all dependent variables can be
def\/ined through the function~$h(r)$, but the equation for $h(r)$
is a fourth-order ordinary dif\/ferential equation. In fact, since
$H\neq 0$, from~(\ref{eq:spvortst}) one obtains that $U\neq 0$.
Hence, $\alpha =\alpha _0$, where $\alpha _0$ is constant. From
the f\/irst and third equations of (\ref{eq:spvortst}), one
f\/inds
\[
\rho =R_0\frac{h^{\prime }}{\sqrt{h^2+1}},\qquad U=\frac{\alpha
_0(h^2+1)}{ h^{\prime }}.
\]
In this case
\[
\dot \rho =-\alpha _0R_0h^{\prime }\left(
\frac{\sqrt{h^2+1}}{h^{\prime }} \right) ^{\prime }
\]
and after substituting $\rho $ and $\dot \rho $ into the formula
for the pressure, one has
\[
p=F(h,h^{\prime },h^{\prime \prime },h^{\prime \prime \prime }),
\]
where the function $F$ is def\/ined by the potential function $W$.
Substituting representations of~$\rho $,~$U$ and~$p$ into the
second equation of (\ref{eq:spvort}), one obtains the fourth-order
ordinary dif\/ferential equation for the function $h(r)$.

\subsection[Invariant solutions of (\ref{eq:spvort}) with $W=-q_0\dot \rho
^2\rho ^{-5/3}+\beta \rho ^{5/3}$]{Invariant solutions of
(\ref{eq:spvort}) with $\boldsymbol{W=-q_0\dot \rho ^2\rho
^{-5/3}+\beta \rho ^{5/3}}$}

System of equations (\ref{eq:spvort}) with the potential function
\[
W=-q_0\dot \rho ^2\rho ^{-5/3}+\beta \rho ^{5/3}
\]
admit the Lie group corresponding to the Lie algebra
$L_3=\{X_0,X_2,X_3\}$.

If $\beta =0$, then there is one more admitted generator $X_1$.
The four-dimensional Lie algebra with the generators
$\{X_0,X_1,X_2,X_3\}$ is denoted by $L_4$.

The structural constants of the Lie algebra $L_4$ are def\/ined by
the table of commutators:
\[
\begin{array}{c|cccc}
& X_0 & X_1 & X_2 & X_3 \\
\hline
X_0 & 0 & X_0 & X_3 & 2X_0 \\
X_1 &  & 0 & X_2 & 0 \\
X_2 &  &  & 0 & -2X_2 \\
X_3 &  &  &  & 0
\end{array}
\]
Solving the Lie equations for the automorphisms, one obtains:
\begin{gather*}
A_0:\left\{
\begin{array}{l}
\widetilde{x}_0  =x_0+a_0(x_1+2x_3)+a_0^2x_2, \\
\widetilde{x}_3  =x_3+a_0x_2,
\end{array}
\right. \qquad A_1:\left\{
\begin{array}{l}
\widetilde{x}_0  =x_0e^{-a_1}, \\
\widetilde{x}_2  =x_2e^{a_1},
\end{array}
\right.
\\
A_2:\left\{
\begin{array}{l}
\widetilde{x}_2  =x_2+a_2(x_1+2x_3)+a_2^2x_0, \\
\widetilde{x}_3  =x_3+a_2x_0,
\end{array}
\right.  \qquad A_3:\left\{
\begin{array}{l}
\widetilde{x}_0  =x_0e^{a_3}, \\
\widetilde{x}_2  =x_2e^{a_3}.
\end{array}
\right.
\end{gather*}
Construction of the optimal system of one-dimensional admitted
subalgebras consists of using the automorphisms $A_i$
$(i=0,1,2,3)$ for simplif\/ications of the coordinates
$(x_0,x_1,x_2,x_3)$ of the generator
\[
X=\sum_{j=0}^3x_jX_j.
\]
Here $k$ is the dimension of the Lie algebra $L_k$  ($k=3,4$). In the case $%
L_3$ one has to assume that the coordinate $x_1=0$.

Beside automorphisms for constructing optimal system of
subalgebras one can use involutions. Equations~(\ref{eq:spvort})
posses the involutions~$E$, corresponding to the change
$t\rightarrow -t$. The involution~$E$ acts on the generator
\[
X=\sum_{j=0}^3x_jX_j.
\]
by transforming the generator $X$ into the generator
$\widetilde{X}$ with the changed coordinates:
\[
E:\left\{
\begin{array}{l}
\widetilde{x}_0  =-x_0, \\
\widetilde{x}_2  =-x_2.
\end{array}
\right.
\]
Here only the changed coordinates are presented.

\subsection{One-dimensional subalgebras}

One can decompose the Lie algebra $L_4$ as $L_4=I\oplus N$, where
$I=L_3$ is an ideal and $N=\{X_1\}$ is a subalgebra of $L_4$.
Classif\/ication of the subalgebra $N=\{X_1\}$ is simple: it
consists of the subalgebras:
\[
N_1=\{0\},\qquad N_2=\{X_1\}.
\]
According to the algorithm \cite{bk:Ovsiannikov[1993opt]} for
construction of an optimal system of one-dimensional sub\-algebras
one has to consider two types of generators:
(a)~$X=x_0X_0+x_2X_2+x_3X_3$, \linebreak
(b)~$X=X_1+x_0X_0+x_2X_2+x_3X_3$. Notice that case (a) corresponds
to the Lie algebra~$L_3$. Hence, classifying the Lie algebra
$L_4$, one also obtains classif\/ication of the Lie algebra $L_3$.

\subsubsection{Case (a)}

Assuming that $x_0\neq 0$, choosing $a_2=-x_3/x_0$, one maps $x_3$
into zero. This means that $\widetilde{x}_3=0$. For simplicity of
explanation, we write it as $x_3(A_2)\rightarrow 0$. In this case
$x_2(A_2)\rightarrow
\widetilde{x}_2=x_2-x_3{}^2/x_0$. If $\widetilde{x}_2\neq 0$, then applying $%
x_2(A_1)\rightarrow \pm 1$, hence, the generator $X$ becomes
\[
X_2+\alpha X_0,\qquad \alpha =\pm 1 .
\]
If $\widetilde{x}_2=0$, then one has the subalgebra: $\{X_0\}.$

In the case $x_0=0$, if $x_3\neq 0$ or $x_2\neq 0$, then, applying
$A_0$, one can obtain $x_0\neq 0$, which leads to the previous
case. Hence, without loss of generality one also assumes that
$x_3=0$, $x_2=0$. Thus, the optimal system of one-dimensional
subalgebras in case (a) consists of the subalgebras
\begin{gather}
\label{optsysL3} \{X_2\pm X_0\},\qquad  \{X_0\}.
\end{gather}
This set of subalgebras also composes an optimal system of
one-dimensional subalgebras of the algebra $L_3$.

\subsubsection{Case (b)}

Assuming that $x_0\neq 0$, choosing $a_2=-x_3/x_0$, one maps $x_3$
into zero. In this case $x_2(A_2)\rightarrow
\widetilde{x}_2=x_2-x_3(1-x_3)/x_0$. If $\widetilde{x}_2\neq 0$,
then applying $A_1$, and $E_2$ (if necessary), one maps the
generator $X$ into
\[
X_1+X_2+\gamma X_0,
\]
where $\gamma \neq 0$ is an arbitrary constant. If
$\widetilde{x}_2=0$, then $x_0(A_0)\rightarrow 0$, and the
generator $X$ becomes~$X_1$.

In the case $x_0=0$, if $2x_3+1\neq 0$ or $x_2\neq 0$, then,
applying $A_0$, one can obtain $x_0\neq 0$, which leads to the
previous case. Hence, without loss of generality one also assumes
that $x_3=-1/2$, $x_2=0$, and the generator $X$ becomes
$X_3-2X_1$.

Thus, the optimal system of one-dimensional subalgebras of the Lie algebra $%
L_4$ consists of the subalgebras
\begin{gather*}
\{X_2\pm X_0\},\qquad \{X_0\},\qquad \{X_1+X_2+\gamma X_0\},\qquad
\{X_3-2X_1\},\qquad \{X_1\},
\end{gather*}
where $\gamma \neq 0$ is an arbitrary constant.

\begin{remark} An optimal system of subalgebras for $W=-q_0 \rho ^{-3}\dot
\rho ^2+\beta \rho ^3$ with arbitrary $\beta $ consists of the
subalgebras (\ref{optsysL3}).
\end{remark}

\begin{remark}
 The subalgebra $\{X_2-X_0\}$ is equivalent to the
subalgebra: $\{X_3\}$.
\end{remark}

\subsection[Invariant solutions of $X_1+X_2+\gamma X_0$]{Invariant solutions of $\boldsymbol{X_1+X_2+\gamma X_0}$}

The generator of this Lie group is
\[
X=\gamma X_0+X_1+X_2=(t^2+t+\gamma )\partial _t+tr\partial
_r-3t\rho
\partial _\rho +(r-U(t+1))\partial _U-\alpha \partial _\alpha .
\]
To f\/ind invariants, one needs to solve the equation
\[
XJ=0,
\]
where $J=J(t,r,\rho ,U,\alpha ,h)$. A solution of this equation
depends on the value of $\gamma $.

Let $\gamma =\mu ^2+1/4$. In this case invariants of the Lie group
are
\begin{gather*}
y=rs,\quad V=s(((t+1/2)^2+\mu ^2)U-rt),\quad R=\rho s^{-3},\quad
\Lambda =\alpha e^{\frac 1\mu \arctan (\frac{2t+1}{2\mu })},\quad
h,
\end{gather*}
where
\[
s=\left( (t+1/2)^2+\mu ^2\right) ^{-1/2}e^{\frac 1{2\mu }\arctan
(\frac{2t+1 }{2\mu })}.
\]
The representation of an invariant solution is
\begin{gather*}
s\left( \left( (t+1/2)^2+\mu ^2\right) U-rt\right) =V(y),\quad
\rho =s^3R(y),\quad \alpha =\Lambda e^{-\frac 1\mu \arctan
(\frac{2t+1}{2\mu } )},\quad h=h(y).
\end{gather*}
Substituting the representation of a solution into
(\ref{eq:spvort}), one obtains the system of four ordinary
dif\/ferential equations
\begin{gather*}
V^{\prime }=-\frac{R^{\prime }}RV+(\Lambda h-8Vy)/(4y^2),\qquad
h^{\prime }=\frac \Lambda V\frac{(h^2+1)}{4y^2},\qquad \Lambda
^{\prime }=\frac \Lambda V,
\\
R^{\prime \prime \prime }=\big(-((8((3(4(44V+5y)y-19\Lambda
h)R+308R^{\prime
}Vy^2)R^{\prime }{}^2 \\
\phantom{R^{\prime \prime \prime }=} {}-3(88R^{\prime
}Vy^2-9\Lambda hR+12(6V+y)Ry)R^{\prime \prime
}R)Vq_0y-9R^{2/3}(4(4(2V+y)V \\
\phantom{R^{\prime \prime \prime }=} {}-(4\mu
^2+1)y^2)y^2+(\Lambda -4hVy)\Lambda
)R^3)y-18(8(R^{2/3}Vy^3+4\Lambda
hq_0)Vy \\
\phantom{R^{\prime \prime \prime }=} {}-(2h^2+1)\Lambda
^2q_0-4(8(5V+y)V-(4\mu ^2+1)y^2)q_0y^2)R^{\prime
}R^2)\big)/(288R^2V^2q_0y^4).
\end{gather*}

Let $\gamma =-\mu ^2+1/4$. A representation of a solution is
\begin{gather*}
s(((t+1/2)^2-\mu ^2)U-rt)=V(y),\qquad \alpha (t+1/2-\mu )^{\frac
1{2\mu
}}(t+1/2+\mu )^{-\frac 1{2\mu }}=\Lambda (y), \\
\rho (t+1/2-\mu )^{3\alpha _1}(t+1/2+\mu )^{3\alpha
_2}=R(y),\qquad h=h(y),
\end{gather*}
where
\begin{gather*}
y=rs,\qquad s=(t+1/2-\mu )^{-\alpha _1}(t+1/2+\mu )^{-\alpha
_2},\;\;\alpha _1= \frac{2\mu -1}{4\mu },\qquad \alpha
_2=\frac{2\mu +1}{4\mu }.
\end{gather*}
In this case
\begin{gather*}
V^{\prime }=-V\frac{R^{\prime }}R+\frac{\Lambda
h-2Vy)}{y^2},\qquad h^{\prime }=\Lambda
\frac{(h^2+1)}{Vy^2},\qquad \Lambda ^{\prime }=\frac \Lambda V,
\\
R^{\prime \prime \prime }=(528R^{\prime \prime }R^{\prime
}RV^2q_0y^4+72R^{\prime \prime }R^2Vq_0y^2(-3\Lambda
h+6Vy+y^2)-616R^{\prime
}\ ^3V^2q_0y^4 \\
\phantom{R^{\prime \prime \prime }=} {}+24R^{\prime
}{}^2RVq_0y^2(19\Lambda h-44Vy-5y^2)+18R^{\prime
}R^2(2R^{2/3}V^2y^4-8\Lambda ^2h^2q_0 \\
\phantom{R^{\prime \prime \prime }=} {}-4\Lambda ^2q_0+32\Lambda
hVq_0y-40V^2q_0y^2-8Vq_0y^3-4\mu ^2q_0y^4+q_0y^4)
\\
\phantom{R^{\prime \prime \prime }=} {}+9R^{2/3}R^3y(4\Lambda
^2-4\Lambda hVy+8V^2y^2+4Vy^3+4\mu ^2y^4-y^4))/(72R^2V^2q_0y^4).
\end{gather*}

Let $\gamma =1/4$. A representation of an invariant solution is
\[
s((t+1/2)^2U-rt)=V(y),\qquad \rho =s^3R(y),\qquad \alpha
=e^{2/(2t+1)}\Lambda (y),\qquad h=h(y),
\]
where
\[
y=rs,\qquad s=\frac 1{(t+1/2)}e^{-1/(2t+1)}.
\]
In this case
\begin{gather*}
V^{\prime }=-V\frac{R^{\prime }}R+\frac{(\Lambda
h-2Vy)}{y^2},\qquad h^{\prime }=\frac \Lambda
V\frac{(h^2+1)}{y^2},\qquad \Lambda ^{\prime }=\frac \Lambda V,
\\
R^{\prime \prime \prime }=(528R^{\prime \prime }R^{\prime
}RV^2q_0y^4+72R^{\prime \prime }R^2Vq_0y^2(-3\Lambda
h+6Vy+y^2)-616R^{\prime
}{}^3V^2q_0y^4 \\
\phantom{R^{\prime \prime \prime }=} {}+24R^{\prime
}{}^2RVq_0y^2(19\Lambda h-44Vy-5y^2)+18R^{\prime
}R^2(2R^{2/3}V^2y^4-8\Lambda ^2h^2q_0-4\Lambda ^2q_0 \\
\phantom{R^{\prime \prime \prime }=} {}+32\Lambda
hVq_0y-40V^2q_0y^2-8Vq_0y^3+q_0y^4)+9R^{2/3}R^3y(4\Lambda
^2-4\Lambda hVy+8V^2y^2 \\
\phantom{R^{\prime \prime \prime }=}
{}+4Vy^3-y^4))/(72R^2V^2q_0y^4).
\end{gather*}

These equations were obtained assuming that $V\neq 0$. The case
$V=0$ leads to
\[
\Lambda = 0,\qquad 2q_0 R^{\prime}- yR^{5/3} = 0.
\]

\subsection[Invariant solutions of $X_3-2X_1$]{Invariant solutions of $\boldsymbol{X_3-2X_1}$}

Invariants of the generator
\[
X_3-2X_1=r\partial _r-3\rho \partial _\rho +U\partial _U+2\alpha
\partial _\alpha
\]
are
\[
U=rV(y),\qquad \rho =r^{-3}R(y),\qquad \alpha =r^2\Lambda
(y),\qquad h=h(y),
\]
where $y=t$. Substitution into equations (\ref{eq:spvort}) gives
that the functions $V(y)$, $R(y)$, $\Lambda (y)$ and~$h(y)$ have
to satisfy the equations
\begin{gather*}
h^{\prime }=\Lambda (h^2+1),\qquad \Lambda ^{\prime }=-2\Lambda
V,\qquad R^{\prime }=\Lambda hR,
\\
3(R^{2/3}+6q_0)(V^{\prime }+V^2)=\Lambda ^2\big(
4q_0(h^2-3)+3(R^{2/3}+6q_0)\big) .
\end{gather*}

\subsection[Invariant solutions of $X_1$]{Invariant solutions of $\boldsymbol{X_1}$}

Invariants of the generator $X_1$%
\[
X_1=t\partial _t-U\partial _U-\alpha \partial _\alpha
\]
are
\[
x,\quad Ut,\quad \rho ,\quad h,\quad \alpha t.
\]
An invariant solution has the representation
\[
U=t^{-1}V(y),\qquad \rho =R(y),\qquad \alpha =t^{-1}\alpha
(y),\qquad h=h(y),
\]
where $y=x$. Substituting into equations (\ref{eq:spvort}), one
obtains
\begin{gather*}
V^{\prime }=-V\frac{R^{\prime }}R+\frac{\Lambda h-2Vy}{y^2},\qquad
h^{\prime }=\frac \Lambda V\frac{(h^2+1)}{y^2},\qquad \Lambda
^{\prime }=\frac \Lambda V,
\\
R^{\prime \prime \prime }=(132R^{\prime \prime }R^{\prime
}RV^2q_0y^4+18R^{\prime \prime }R^2Vq_0y^2(-3\alpha
h+6Vy+y^2)-154R^{\prime
}{}^3V^2q_0y^4 \\
\phantom{R^{\prime \prime \prime }=}{}+6R^{\prime
}{}^2RVq_0y^2(19\alpha h-44Vy-5y^2)+9R^{\prime
}R^2(R^{2/3}V^2y^4-4\alpha ^2h^2q_0-2\alpha ^2q_0 \\
\phantom{R^{\prime \prime \prime }=}{} +16\alpha hVq_0y
-20V^2q_0y^2-4Vq_0y^3)+9R^{2/3}R^3y(\alpha ^2-\alpha
hVy\\
\phantom{R^{\prime \prime \prime
}=}{}+2V^2y^2+Vy^3))/(18R^2V^2q_0y^4).
\end{gather*}
Here it is assumed that $V\neq 0$. The case $V=0$ only leads to
the condition $\Lambda = 0$.

\subsection[Invariant solutions of $X_2+X_0$]{Invariant solutions of $\boldsymbol{X_2+X_0}$}
\[
X_2=t(t\partial _t+r\partial _r-U\partial _U-3\rho \partial _\rho
)+r\partial _U.
\]
Invariants of the generator
\[
X_2+X_0=(t^2+1)\partial _t+tr\partial _r-3t\rho \partial _\rho
+(r-tU)\partial _U
\]
are
\[
r(t^2+1)^{-1/2},\quad U(t^2+1)^{1/2}-rt(t^2+1)^{-1/2},\quad \rho
(t^2+1)^{3/2},\quad \alpha ,\quad h.
\]
An invariant solution has the representation
\[
U(t^2+1)^{1/2}-rt(t^2+1)^{-1/2}=V(y),\!\qquad \rho
=(t^2+1)^{-3/2}R(y),\!\qquad \alpha =\alpha (y),\!\qquad h=h(y).
\]
where $y=r(t^2+1)^{-1/2}$. Substituting into equations
(\ref{eq:spvort}), one has to study two cases: (a) $V=0$, and (b)
$V\neq 0$.

Assuming $V=0$, one obtains that $\Lambda = 0$, and the function
$R$ satisf\/ies the equation
\[
2(5\beta R^{4/3} - 9q_0)R^{\prime}+ 9yR^{5/3} = 0.
\]

If $V\neq 0$, then one obtains
\begin{gather*}
V^{\prime }=-V\frac{R^{\prime }}R+\frac{(\Lambda
h-2Vy)}{y^2},\qquad h^{\prime }=\frac \Lambda
V\frac{(h^2+1)}{y^2},\qquad \Lambda ^{\prime }=0,
\\
R^{\prime \prime \prime }=(132R^{\prime \prime }R^{\prime
}RV^2q_0y^4+54R^{\prime \prime }R^2Vq_0y^2(-\Lambda h+2Vy)-154R^{\prime }{}^3V^2q_0y^4 \\
\phantom{R^{\prime \prime \prime }=}{} +6R^{\prime
}{}^2RVq_0y^2(19\Lambda h-44Vy)-10R^{1/3}R^{\prime }R^3\beta
y^4+9R^{\prime }R^2(R^{2/3}V^2y^4-4\Lambda ^2h^2q_0 \\
\phantom{R^{\prime \prime \prime }=}{}-2\Lambda ^2q_0+16\Lambda
hVq_0y-20V^2q_0y^2+2q_0y^4)+9R^{2/3}R^3y(\Lambda
^2-\Lambda hVy+2V^2y^2 \\
\phantom{R^{\prime \prime \prime }=}{}-y^4))/(18R^2V^2q_0y^4).
\end{gather*}

\subsection[Invariant solutions of $X_2-X_0$]{Invariant solutions of $\boldsymbol{X_2-X_0}$}

Since the Lie algebra $\{X_2-X_0\}$ is equivalent to the Lie
algebra with the generator $\{X_3\}$, then for the sake of
simplicity an invariant solution with respect to
\[
X_3=2t\partial _t+r\partial _r-U\partial _U-3\rho \partial _\rho
\]
is considered here. Invariants of the generator $X_3$ are
\[
rt^{-1/2},\quad Ut^{1/2},\quad \rho t^{3/2},\quad h,\quad \alpha .
\]
An invariant solution has the representation{\samepage
\[
U=t^{-1/2}V(y),\qquad \rho =t^{-3/2}R(y),\qquad \alpha =\alpha
(y),\qquad h=h(y),
\]
where $y=rt^{-1/2}$.}

Substituting into equations (\ref{eq:spvort}), one has to study
two cases: (a) $V-y/2=0$, and \linebreak \mbox{(b)~$V-y/2\neq 0$}.

Assuming $V-y/2=0$, one obtains that $\Lambda = 0$, and the
function $R$ satisf\/ies the equation
\[
2(20\beta R^{4/3}+ 9q_0)R^{\prime}- 9yR^{5/3} = 0.
\]

If $V-y/2\neq 0$, then one obtains
\begin{gather*}
V^{\prime }=(y/2-V)\frac{R^{\prime }}R+\frac{2\Lambda
h-(4V-3y)y}{2y^2} ,\qquad h^{\prime }=\frac \Lambda
{(V-y/2)}\frac{(h^2+1)}{y^2},\qquad \Lambda ^{\prime }=0,
\\ 
R^{\prime \prime \prime }=  (2(y-2V)^2q_0 y^3(66R^{\prime \prime }R^{\prime
}Ry+54R^{\prime \prime }R^2-77R^{\prime }{}^3y-132R^{\prime }{}^2R) \\
\phantom{R^{\prime \prime \prime }=}{} +9(y-2V)^2y^2R^2(R^{\prime }R^{2/3}y^2-20R^{\prime
}q_0+2R^{5/3}y)-18R^{\prime }R^2y^4q_0 \\
\phantom{R^{\prime \prime \prime }=}{} +6(y-2V)\alpha hy(18R^{\prime \prime }R^2q_0y-38R^{\prime
}{}^2Rq_0y-48R^{\prime }R^2q_0+3R^{2/3}R^3y) \\
\phantom{R^{\prime \prime \prime }=}{} -72R^{\prime }\alpha ^2R^2q_0(2h^2+1)-40R^{10/3}R^{\prime }\beta
y^4+36R^{2/3}\alpha ^2R^3y \\
\phantom{R^{\prime \prime \prime }=}{} +9R^{11/3}y^5)/(18R^2q_0 y^4(y-2V)^2).
\end{gather*}

\section{Conclusion}

In this paper the complete group classif\/ication of the
three-dimensional equations describing a~motion of f\/luids with
internal inertia (\ref{tab:grclass}) is given. The
classif\/ication is considered with respect to the potential function $W(\rho,\dot\rho)$. Detailed study of one class
of partially invariant solutions (the Ovsyannikov vortex) for a
particular potential function is presented. This solution is
essentially three-dimensional.

\subsection*{Acknowledgments}

The work of P.S.\ has been supported by scholarship of the
Ministry of University Af\/fairs of Thailand. The authors also
thank S.L.~Gavrilyuk for fruitful discussions, and  E.~Schulz for
his kind help.

\pdfbookmark[1]{References}{ref}
\LastPageEnding

\end{document}